\documentclass[twocolumn,aps]{revtex4-2}
\usepackage[latin9]{inputenc}
\usepackage{array}
\usepackage{multirow}
\usepackage{amsmath}
\usepackage{amssymb}
\usepackage{graphicx}

\makeatletter

\providecommand{\tabularnewline}{\\}

\@ifundefined{textcolor}{}{%
\definecolor{BLACK}{gray}{0}
\definecolor{WHITE}{gray}{1}
\definecolor{RED}{rgb}{1,0,0}
\definecolor{GREEN}{rgb}{0,1,0}
\definecolor{BLUE}{rgb}{0,0,1}
\definecolor{CYAN}{cmyk}{1,0,0,0}
\definecolor{MAGENTA}{cmyk}{0,1,0,0}
\definecolor{YELLOW}{cmyk}{0,0,1,0}
}

\usepackage{upgreek}
\usepackage{braket}

\providecommand{\tabularnewline}{\\}

\@ifundefined{textcolor}{}{%
\definecolor{BLACK}{gray}{0}
\definecolor{WHITE}{gray}{1}
\definecolor{RED}{rgb}{1,0,0}
\definecolor{GREEN}{rgb}{0,1,0}
\definecolor{BLUE}{rgb}{0,0,1}
\definecolor{CYAN}{cmyk}{1,0,0,0}
\definecolor{MAGENTA}{cmyk}{0,1,0,0}
\definecolor{YELLOW}{cmyk}{0,0,1,0}
}

\usepackage{color}

\def\ket#1{|#1\rangle}

\def\NOT(#1,#2){\OneQubitGate(#1,#2){$X$}}

\@ifundefined{showcaptionsetup}{}{%
\PassOptionsToPackage{caption=false}{subfig}}

\makeatother

\begin{document}
\title{Optimization of a quantum control sequence for initializing an NV
spin register}
\author{T. Chakraborty$^{1,2}$, J. Zhang$^{1}$ and D. Suter$^{1,3}$}
\affiliation{$^{1}$Fakult{ä}t Physik, Technische Universit{ä}t Dortmund, D-44221
Dortmund, Germany}
\affiliation{$^{2}$QuTech and Kavli Institute of Nanoscience, Delft University
of Technology, 2600 GA Delft, The Netherlands}
\affiliation{$^{3}$Department of Physics and NMR Research Center, Indian Institute
of Science Education and Research, Pune 411008, India}
\begin{abstract}
Implementation of many quantum information protocols require an efficient
initialization of the quantum register. In the present report, we
optimize a population trapping protocol for initializing a hybrid
spin register associated a single nitrogen vacancy (NV) center in
diamond. We initialize the quantum register by polarizing the electronic
and the nuclear spins of the NV with a sequence of microwave, radio-frequency
and optical pulses. We use a rate equation model to explain the distribution
of population under the effect of the optical pulses. The model is
compared to the experimental data obtained by performing partial quantum
state tomography. To further increase the spin polarisation, we propose
a recursive protocol with optimized optical pulses.
\end{abstract}
\maketitle

\section*{I. Introduction}

A nitrogen atom replacing one carbon atom of a diamond lattice and
an adjacent carbon vacancy site form a nitrogen vacancy (NV) center
in diamond, which is an optically active atomic defect center \citep{Jelezko_review,suter2017single}.
The NV center, having excellent room temperature (RT) quantum properties
like long spin coherence time \citep{balasubramanian2009ultralong},
offers a stable spin-photon interface \citep{kubanek2012quantum,bernien2012two},
coherent optical transitions and so on, which have been successfully
applied in a number of aspects of the emerging field of quantum technology
\citep{pezzagna2021quantum,childress2013diamond,fabbri2020quantum}.
For instance, NV centers were used to demonstrate spin-photon entanglement
with scalability \citep{vasconcelos2020scalable}, quantum teleportation
\citep{pfaff2014unconditional}, efficient implementation of quantum
algorithms \citep{zhang2020efficient}, entanglement distribution
over a multinode quantum network \citep{pompili2021realization},
photonic quantum repeater \citep{nemoto2016photonic}, quantum sensing
platforms with remarkable sensitivity \citep{degen2017quantum, schirhagl2014nitrogen}
and so on.

Having a large band gap ($5.4$ eV), diamond has an unoccupied conduction
band which does not allow interaction of free electrons with the NV
centers \citep{larsson2008electronic}. Because of its high Debye
temperature ($\approx1800$ $^{0}$C), the interaction with phonons
is weak at RT \citep{doherty2012theory}. In addition, naturally abundant
diamond crystals contain magnetic impurities of very low concentration
($1.1\%$ $^{13}$C) and growth of ultrapure single crystal diamond
contain NVs with slowly dephasing spins \citep{chakraborty2019cvd}.
Thus diamond, as a host, ensures that NV centers can be considered
as isolated quantum systems, where the long lived electronic and nuclear
spins associated with NV can act as a quantum register \citep{Dutta_Science}.
Several experiments demonstrated local control of the NV spins with
high fidelity through initialization, coherent manipulation and read
out using optical, microwave (MW), and radio frequency (RF) pulses
\citep{Dutta_Science,childress2006coherent,gaebel2006room,chakraborty2017polarizing}.

Often for quantum operations, hybrid spin register associated with
NVs are exploited \citep{Dutta_Science,APL(105)242402,chakraborty2017polarizing,rao2020optimal}.
Commonly, these registers include vacancy electronic spins being coupled
to the nuclear spins from $^{14}$N or neighboring $^{13}$C atoms
\citep{suter2017single}. For efficient implementation of several
NV-based quantum information protocols, it is necessary that the initial
state of the spin register has high purity. Hence, one needs to optimize
methods which simultaneously polarize the electronic and nuclear spins
of the NV center. Although it is straightforward to achieve almost
complete polarization of the electronic spins through optical pumping,
it does not allow polarization of the nuclear spins. Due to having
an ultra long spin coherence time, nuclear spins are important candidates
for storing and processing quantum information. Hence, to use nuclear
spin as a resource in quantum protocols, it is necessary to perform
the nuclear polarization deterministically. In this context a number
of techniques are reported. Polarization of $^{15}$N nuclear spin
is performed by accessing the excited state level anti-crossing for
a single NV at RT \citep{jacques2009dynamic}. By means of transferring
polarization from electronic spins, hyperpolarization of $^{13}$C
nuclear spin ensembles is achieved using a sequence of laser and microwave
pulses, and by exploiting the fact that the nuclear spin eigenstates
do not have same quantization axes in different electron spin manifolds,
which occurs because NV electron spin S=1.\citep{alvarez2015local}.
Moreover, bulk nuclear polarization for NV centers are achieved by
applying laser induced dynamic nuclear polarization (DNP) schemes
\citep{scheuer2016optically,king2015room}.

In the present case, we initialize the spin register through polarizing
the electronic and nuclear spins by applying a sequence of laser,
MW and RF pulses. We consider a hybrid two qutrit system where the
vacancy electron is coupled to the $^{14}$N nuclear spin, and perform
the spin manipulation in a subspace of the spins. Such a protocol
for initialization of the NV spins has been demonstrated earlier \citep{APL(105)242402,chakraborty2017polarizing,rao2020optimal}.
However, In the present case we report a detailed optimization of
this sequence, specifically the optical pumping procedure. Employing
optically detected magnetic resonance (ODMR) spectroscopy techniques
we perform partial state tomography and investigate the population
dynamics of the relevant quantum states using a rate equation model.
We have performed an additional simulation to test if the present
sequence, when executed in an iterative way with optimized laser pulses,
can enhance the polarization further. Thus, our optimized sequence
with known optical pumping parameter can perform NV spin manipulation
in a deterministic way.

The paper is organized as the following. First, we describe the system
and the initialization sequence. Then we present the experimental
results and demonstrate its comparison with the solution of the formulated
rate equation model. Finally, we present the results of the multi-cycle
simulation.

\section*{II. System and polarization procedure}

\begin{figure}
\includegraphics[width=1\columnwidth]{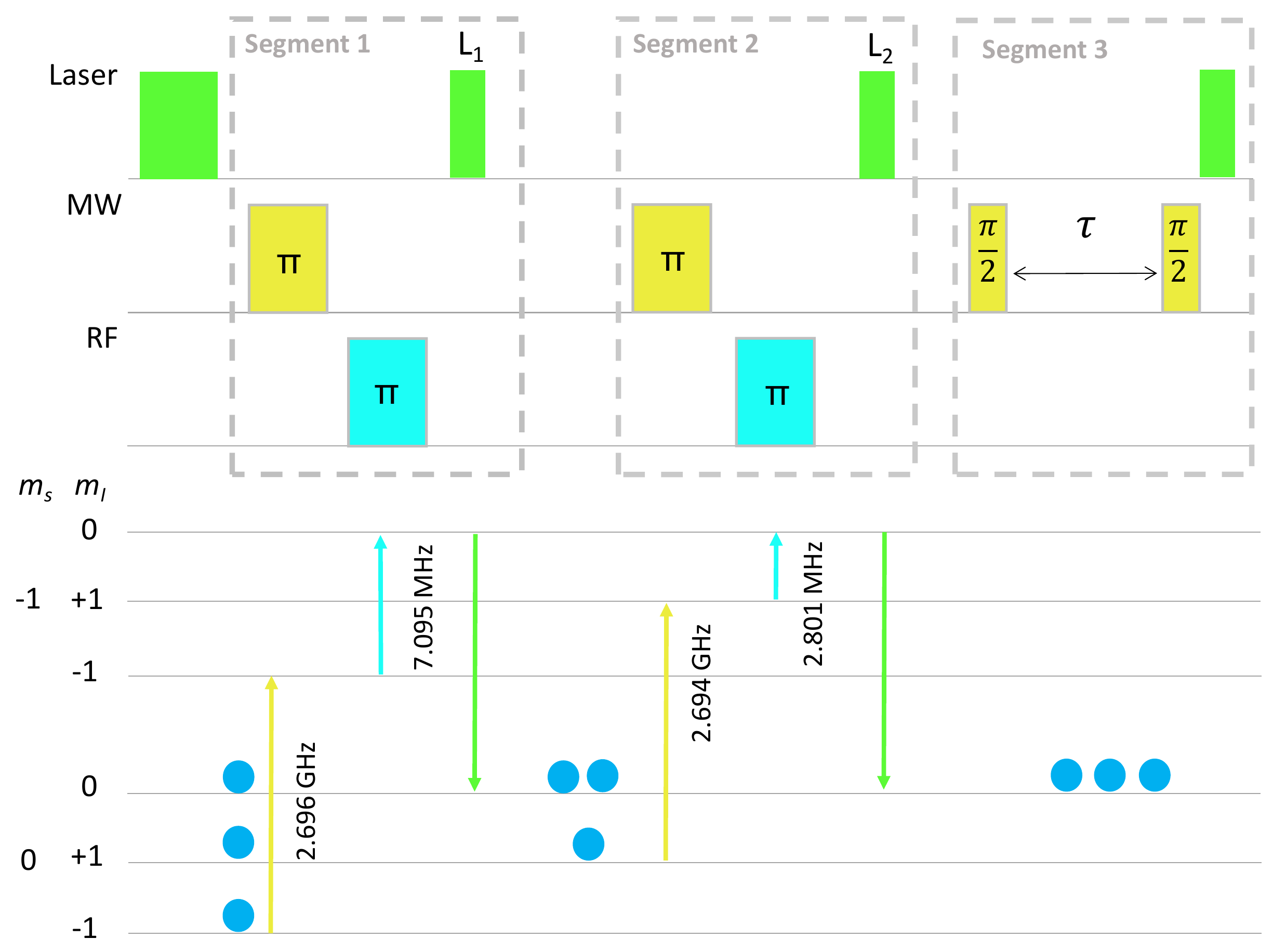}\caption{Sequence of microwave (MW), radio frequency (RF) and $532$ nm laser
pulses used for initialization. The sequence contains three parts:
segment $1$ and $2$ for addressing the nuclear spins $m_{I}=-1$
and $+1$, and segment $3$ measures free induction decay of the electronic
spin. The energy levels with the relevant transitions are shown where
the population of the states are represented by filled circles. \label{fig:pulse_seq}}
\end{figure}

The experiments were performed in a home-built confocal microscope
setup integrated with an electronic circuit which generates the necessary
MW and RF signals for controlling the electronic and nuclear spins,
as described in ref. \citep{chakraborty2017polarizing}. We chose
the vacancy electronic spin $S=1$ coupled with the $^{14}$N nuclear
spin $I=1$ of an isolated NV center embedded in a $99.998\%$ $^{12}$C
enriched diamond as the quantum register for our experiments. The
system Hamiltonian, when the field is aligned along the NV axis, is

\begin{eqnarray}
H & = & DS_{z}^{2}-{\gamma}_{e}BS_{z}+PI_{z}^{2}-{\gamma}_{n}BI_{z}+AS_{z}I_{z}.\label{eq:1}
\end{eqnarray}

$S_{z}$ and $I_{z}$ are the $z$-components of the electronic and
nuclear spins, the zero field splitting is $D=2.87$ GHz, electronic
and nuclear gyromagnetic ratios are ${\gamma}_{e}=-28$ GHzT$^{-1}$
and ${\gamma}_{n}=-3.1$ MHzT$^{-1}$, respectively, the nuclear quadrapole
coupling $P=4.5$ MHz and the hyperfine coupling $A=-2.16$ MHz \citep{chakraborty2017polarizing}.
We performed the experiments in a magnetic field $B=6.1$ mT along
the symmetry axis of the NV.

Our initialization procedure relies on a population trapping (PT)
protocol where by applying a number of $\pi$ pulses we sequentially
accumulate the population into the $m_{I}=0$ state. For this purpose,
we perform the spin manipulation in a subspace spanned by $m_{s}=0,-1$
and $m_{I}=0,\pm1$. We write the states in the notation $\ket{m_{s},m_{I}}$
in this paper. A coherent population trapping (CPT) protocol couples
an initial and a target state to an intermediate excited state by
applying coherent electromagnetic fields, which trap the population
in the superposition of the initial and the target states \citep{jamonneau2016coherent,nicolas2018coherent}.
However, in our case, instead of applying coherent driving fields
simultaneously, we sequentially apply MW, RF and optical pulses to
drive the population between the initial state $\ket{0,\pm1}$ and
the target state $\ket{0,0}$ through certain intermediate states.
Therefore, we term this method a ``population trapping'' protocol.
Fig.\ref{fig:pulse_seq} schematically represents the PT pulse sequence.
We illustrate the PT protocol later in this section. The relevant
energy eigenstates of the six level system are shown in Fig.\ref{fig:pulse_seq}.
The present sequence manipulates the spins associated with the NV$^{-}$
charge state. The transition frequencies and Rabi frequencies for
the electronic and nuclear transitions, which we address in the sequence,
are shown in table \ref{tab:transitions}. In the beginning, a $5$$\mu$s
long laser pulse initializes the system into the bright ($m_{s}=0$)
state of the electron spin while the nuclear spin is fully depolarized.
Now we aim to optimally polarize both the electron and the nuclear
spins by applying a combination of MW, RF and laser pulses. We perform
this task in two segments: through segment(seg)1 and seg2, we address
the population associated with the $m_{I}=-1$ and $+1$ states in
turn. Seg3 contains the pulses which we use for analysing the resulting
state. Seg1 includes a MW and an RF $\pi$ pulse which drive the transitions
$\ket{0,-1}$ $\leftrightarrow$ $\ket{-1,-1}$ and $\ket{-1,-1}$
$\leftrightarrow$ $\ket{-1,0}$. Although these pulses polarize the
nuclear spin, the electron spin becomes depolarized. To repolarize
the electronic spin, we apply a second laser pulse. The effect of
this laser pulse is primarily a polarisation of the electron spin,
but it also partly depolarizes the $^{14}$N nuclear spin \citep{chakraborty2017polarizing}.
We therefore adjust the duration of the laser pulse to maximize the
electronic and the nuclear spin polarization. This task is performed
by measuring the influence of the laser pulse duration on the population
transfer dynamics and by interpreting the results using a rate equation
model, which is discussed in the subsequent section.

\begin{table}
\caption{\label{tab:transitions} Relevant transition frequencies and Rabi
frequencies, that we use in our initialization sequence}

\begin{ruledtabular}
\begin{tabular}{ccc}
Transitions  & Transition frequency  & Rabi frequency \tabularnewline
$\ket{0,-1}$ $\leftrightarrow$ $\ket{-1,-1}$  & $2.696$ GHz & $8.3$ MHz\tabularnewline
$\ket{0,+1}$ $\leftrightarrow$ $\ket{-1,+1}$  & $2.694$ GHz & $8.3$ MHz\tabularnewline
$\ket{-1,-1}$ $\leftrightarrow$ $\ket{-1,0}$  & $2.801$ MHz & $3.87$ kHz\tabularnewline
$\ket{-1,+1}$ $\leftrightarrow$ $\ket{-1,0}$  & $7.095$ MHz & $3.55$ kHz\tabularnewline
\end{tabular}\label{table transitions}
\end{ruledtabular}

\end{table}

Seg2 transfers the population from $\ket{0,+1}$ to $\ket{-1,0}$
using a MW and an RF $\pi$ pulse which act selectively on the transitions
$\ket{0,+1}$ $\leftrightarrow$ $\ket{-1,+1}$ and $\ket{-1,+1}$
$\leftrightarrow$ $\ket{-1,0}$, and swap the populations between
them. Next, in close analogy to seg1, we repolarize the electronic
spin with a laser pulse and optimize its duration. A proper optimization
of the control pulses maximises the population of the $\ket{00}$
state. The optimization of the laser pulses and the experimental analysis
of the final populations of the relevant states are described in the
next section.

\section*{III. Experimental results and comparison with the rate equation model}

\begin{figure}
\includegraphics[width=1\columnwidth]{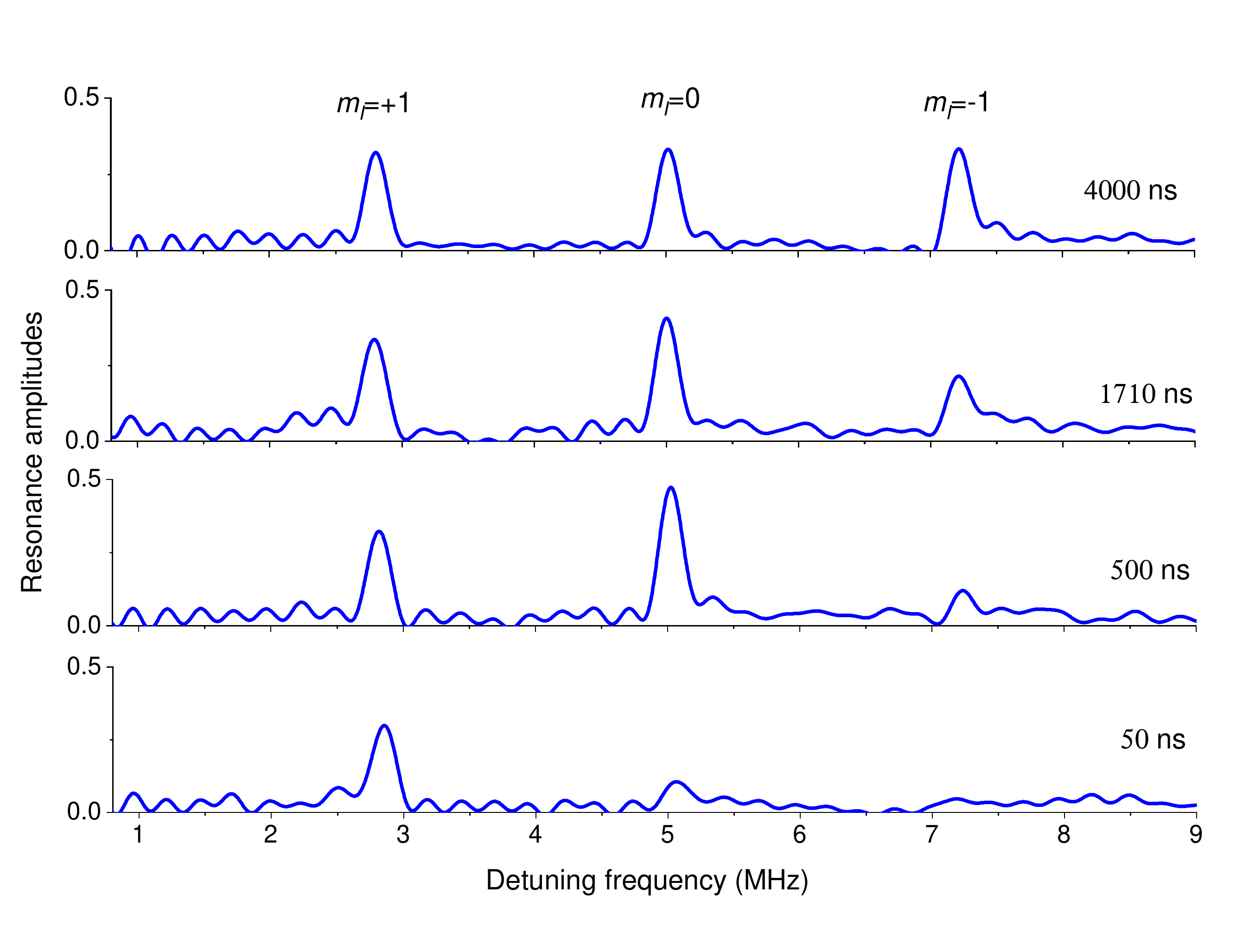} \caption{Electron spin resonance spectra obtained by Fourier transforming the
free induction decay data for the electronic spin. The four spectra
corresponds to four different durations of the laser pulse L$_{1}$
in seg1. The amplitudes of the three lines in the spectra are proportional
to the populations of the three nuclear spin states as marked in the
figure. \label{fig:Spectra_seg1}}
\end{figure}

\begin{figure}
\includegraphics[width=1\columnwidth]{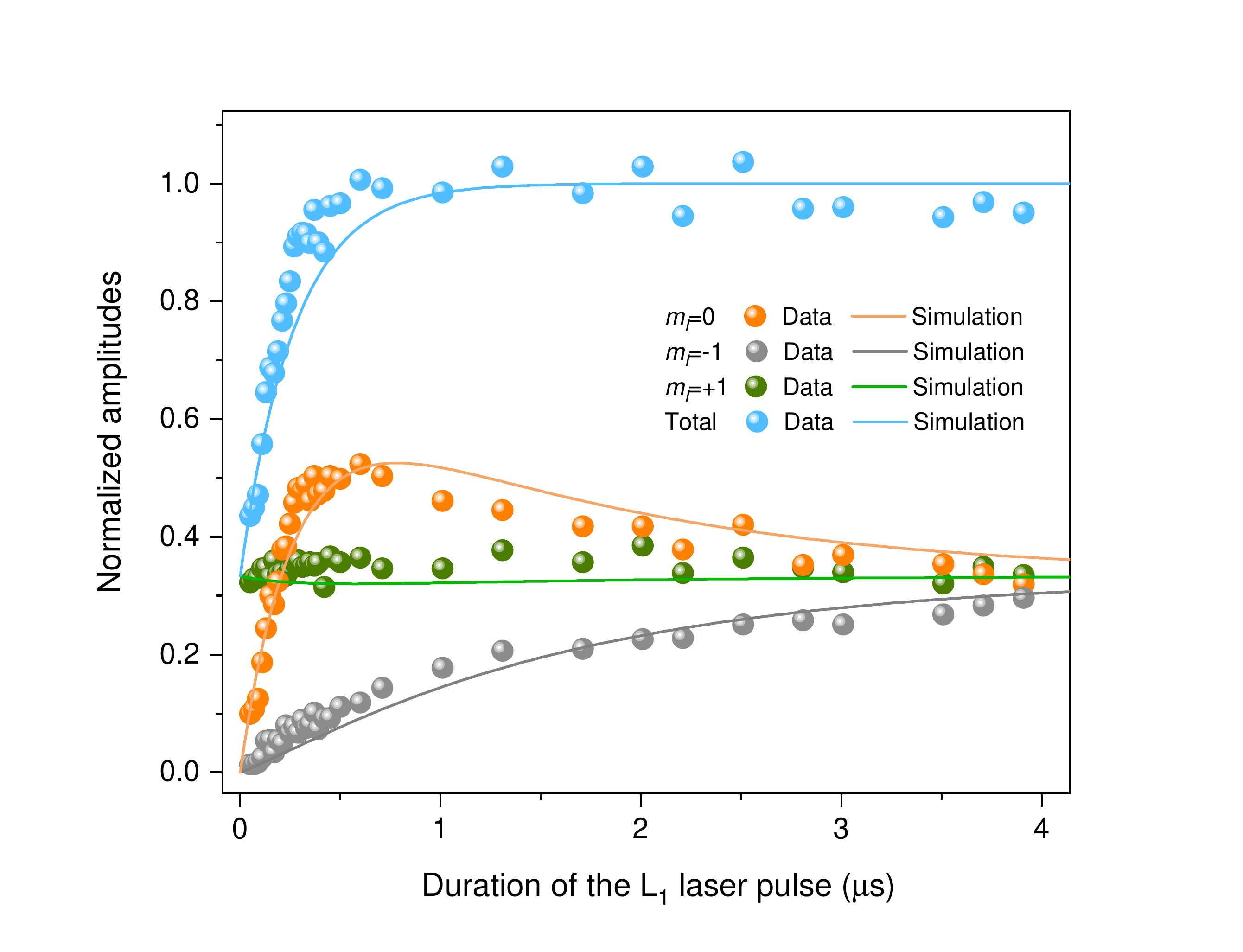}\caption{The variation of spectral amplitudes that correspond to the populations
of the nuclear spin states $m_{I}=0,\pm1$ and their sum as a function
of the duration of the laser pulse L$_{1}$ in seg1. The solid lines
represent the simulation using the rate equation model. We used $1/k_{S}=0.27$
$\mu$s and $1/k_{I}=4.76$ $\mu$s \citep{chakraborty2017polarizing}.
\label{fig:populations_seg1}}
\end{figure}

We implement the proposed initialization protocol and perform the
optimization of the L$_{1}$ and L$_{2}$ laser pulses in two steps.
After polarizing the electronic spin using the $5$$\mu$s laser pulse,
we test and optimize seg1, and analyze the resulting populations of
the states through the measurement protocol of seg3. During this process,
we do not implement seg2. To investigate the redistribution of population
under the influence of the laser pulse L$_{1}$, we run seg1 for different
values of the laser pulse duration, and perform partial state tomography.
This is done by capturing Ramsey type free induction decays (FIDs)
of the electronic spin using the control sequence in seg3. In seg3,
the FID sequence contains two ${\pi}/2$ MW pulses which are separated
by a free evolution time $\tau$ and are resonant between the $m_{s}=0$
and $-1$ levels. The first ${\pi}/2$ pulse creates a coherent superposition
of the $\ket{m_{s}=0}$ and $\ket{m_{s}=-1}$ states, which evolves
for a time $\tau$. The coherence is then converted into population
difference by the second ${\pi}/2$ pulse and read out using a laser
pulse of $300$ ns duration. We apply hard (high power) ${\pi}/2$
pulses such that all the three nuclear spin-conserving transitions
between $m_{s}=0$ and $-1$ are excited. A Fourier transform of the
time domain FID data allows us to capture spectral amplitudes associated
with the $\ket{m_{I}=0,\pm1}$ states in the frequency domain. Thus
the spectra contain three lines whose amplitudes $A_{m_{I}}$ are
proportional to the difference of populations between the $m_{S}=0$
and $-1$ states: $A_{m_{I}}=P_{\ket{0,m_{I}}}-P_{\ket{-1,m_{I}}}$.

To explain the laser induced population transfer dynamics among the
$6$ energy levels of the spin register, we formulate a rate equation
model. We define a column matrix representing the populations of the
six states and call it the population vector $\vec{P}$. We express
the rate equation in the following form:

\begin{eqnarray}
\frac{d\vec{P}}{dt} & = & M(k_{s},k_{I})\vec{P}.\label{eq:2}
\end{eqnarray}

Since the spin manipulation is performed in the subspace spanned by
the $\ket{m_{S}=0,-1}\otimes\ket{m_{I}=0,\pm1}$ states, $m_{S}=+1$
state remains unoccupied throughout the spin control sequence. The
populations of the states in $\vec{P}$ follow the order $(m_{s},m_{I})=(0,-1;0,+1;0,0;-1,-1;-1,+1;-1,0)$
which is the same order for the energy levels shown in Fig.\ref{fig:pulse_seq}.
The rate matrix $M$ describes the transfer of population between
the states due to the effect of the laser pulse and is given in the
appendix. We assume that the electronic spin changes from $m_{S}=-1$
to $0$ with a rate of $k_{s}$ and the nuclear spin changes between
any of the states $\ket{m_{I}=0,\pm1}$ at a rate $k_{I}$. Assuming
that the sum of the populations is unity, the spectral amplitudes
for the $^{14}$N nuclear spins $m_{I}=0$ and $\pm1$ are normalized
with reference to the initial state amplitudes experimentally measured
after applying the $5$$\mu$s long laser pulse, which equally populates
all nuclear spin sublevels. Thus, the normalized population vector
for this state is $\vec{P}=\frac{1}{3}(1,1,1,0,0,0)$. The MW and
RF pulses of seg1 ideally convert this into $\vec{P}=\frac{1}{3}(0,1,1,0,0,1)$
, which is the initial condition for the dynamics during the laser
pulse. The solution of Eq.\ref{eq:2} for this initial condition is
given in the appendix.

Fig.\ref{fig:Spectra_seg1} exhibits $4$ experimental spectra measured
after the implementation of seg1 with durations of $50$, $500$,
$1700$ and $4000$ ns of the laser pulse L$_{1}$. The spectra consist
of three lines associated with the nuclear spins states $m_{I}=0$
and $\pm1$ which are separated by the NV hyperfine interaction. The
MW and RF pulses in seg1 do not control the population of the $\ket{0,+1}$
state. Thus it maintains its initial population of $1/3$ after seg1.
For the shortest duration of the L$_{1}$ pulse ($50$ ns), the electron
spin does not repolarize significantly. Accordingly, the $m_{I}=0$
states in the $m_{s}=0$ and $-1$ subspaces have almost equal population,
resulting in a very small amplitude $A_{0}$ of the $m_{I}=0$ line
in the spectrum. However, with increasing the duration of the L$_{1}$
pulse, the population of the $\ket{m_{I}=0,m_{s}=0}$ state increases,
as the nuclear spin is mostly unaffected while the electronic spin
is transferred from $m_{S}=-1$ to $m_{S}=0$. If the laser pulse
duration is increased further, the nuclear spin becomes depolarized.
Fig.\ref{fig:Spectra_seg1} shows that when the duration of L$_{1}$
is $4$$\mu$s, the three lines have equal amplitudes.

The evolution of the amplitudes $A_{0}$ and $A_{\pm1}$ and the total
population of the three nuclear spin states in the $m_{S}=0$ subspace
are shown in Fig.\ref{fig:populations_seg1}. The amplitudes increase
with increasing pulse duration. $A_{0}$ reaches a maximum when the
duration of the L$_{1}$ pulse is $500$ ns. We optimize the laser
pulse duration such that we obtain the maximum population of the $\ket{0,0}$
state. This optimized pulse allows us to obtain the maximum polarization
of the electron spin with the minimum loss of the nuclear spin polarization.
When the duration of the L$_{1}$ pulse reaches $4$$\mu$s, the total
population of the $m_{S}=0$ state reaches unity which signifies complete
electronic polarization. $A_{0}$ and $A_{-1}$ converge to $1/3$
which indicates complete nuclear spin depolarization. $A_{+1}$ maintains
a constant value of $1/3$ for all durations of the laser pulse. The
amplitudes simulated using our rate equation model are in excellent
agreement with the experimental data. Under our experimental conditions,
$k_{s}$ and $k_{I}$ are similar to the values reported in ref. \citep{chakraborty2017polarizing}.

Next, we set the duration of the L$_{1}$ pulse to $500$ ns and proceed
with the experiments using the sequence of pulses contained in seg2.
In the beginning of seg2, we start with the populations $\vec{P}=(0.07,0.33,0.55,0,0,0.05)$.
Since we address the $m_{I}=+1$ state in this stage, the MW and RF
pulses create the population vector $\vec{P}=(0.07,0,0.55,0,0.05,0.33)$.
Using this as the initial condition we solve the rate equation model.
The solution gives analytical expression of the population of the
six levels as a function of the laser pulse duration which is shown
in the appendix.

\begin{figure}
\includegraphics[width=1\columnwidth]{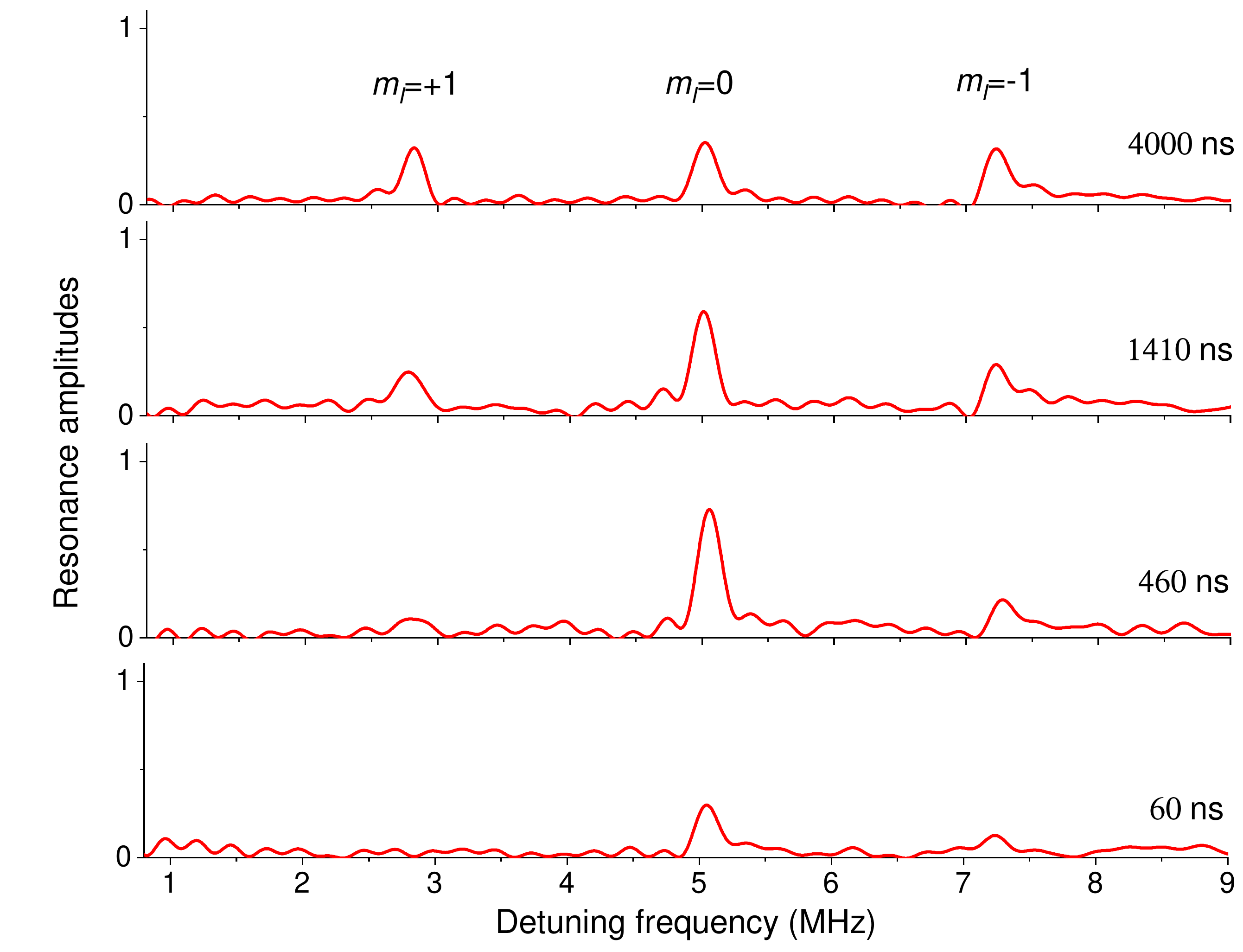}\caption{Fourier transform of free induction decay of the electronic spin for
four different durations of L$_{2}$. The spectra show three peaks
for the three nuclear spin states $m_{I}=0$ and $\pm1$ of $^{14}$N.
\label{fig:Spectra_seg2}}
\end{figure}

\begin{figure}
\includegraphics[width=1\columnwidth]{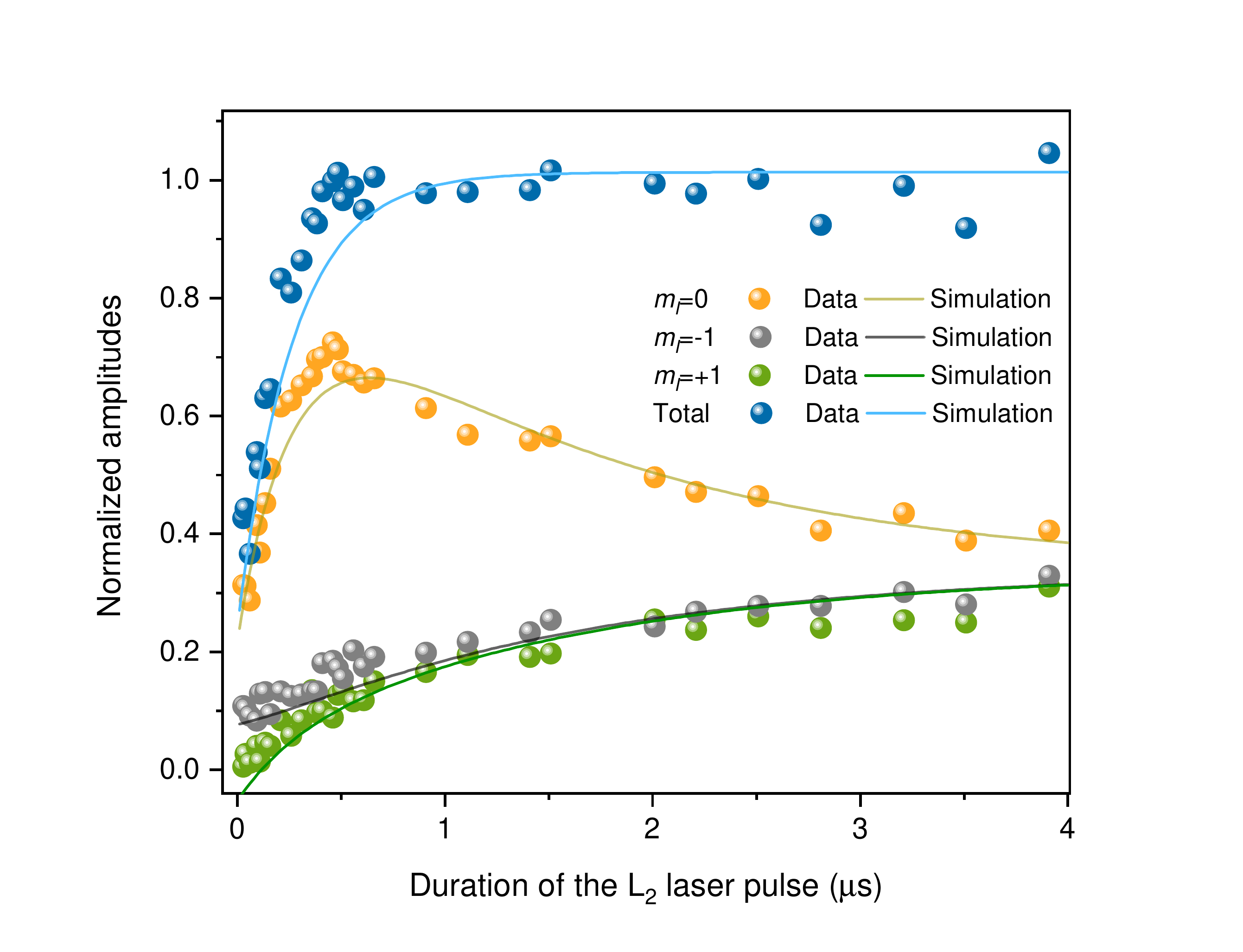}\caption{Amplitudes of the three resonance lines corresponding to the populations
of $m_{I}=0,\pm1$ as a function of the laser pulse (L$_{2}$) duration.
The simulation performed using the rate equation model (with $1/k_{S}=0.27$
$\mu$s and $1/k_{I}=4.76$ $\mu$s \citep{chakraborty2017polarizing})
are shown by the solid curves. \label{fig:populations_seg2}}
\end{figure}

Four spectra measured for $60$, $460$, $1410$ and $4000$ ns duration
of the laser pulse L$_{2}$ are shown in Fig.\ref{fig:Spectra_seg2}.
Similar to the earlier case, due to electron spin polarization, an
increase in pulse duration increases the population difference between
the states $\ket{0,0}$ and $\ket{-1,0}$. $A_{0}$ reaches its maximum
for a pulse duration of $460$ ns. The evolution of the amplitudes
with the laser pulse duration can better be observed in Fig. \ref{fig:populations_seg2}.
The nuclear spin starts to depolarize, populating the $m_{I}$ $=\pm1$
states as the laser pulse duration becomes longer and eventually the
nuclear spin depolarizes completely . The rate equation model is well
consistent with experimental data as it can be seen in Fig. \ref{fig:populations_seg2}.

\section*{IV. Iteration to enhance purity}

Earlier experimental studies of population transfer dynamics \citep{chakraborty2017polarizing,rao2020optimal}have
shown that laser induced depolarization of the nuclear spin limits
the overall purity. An important contribution to loosing nuclear spin
polarization is the charge state conversion dynamics of the NV center.
The two charge states have different hyperfine interactions. Thus,
switching from one charge state to another can result in nuclear depolarization
\citep{chakraborty2017polarizing}. This prediction is supported by
recent results showing that the NV$^{-}$ to NV$^{0}$ conversion
rate is reduced if light $594$ nm light is used instead of 532 nm
\citep{rao2020optimal}. In this section we show that it is possible
to enhance the polarization of the hybrid spin register beyond the
values obtained for a single cycle in section III, when MW, RF and
properly optimized $532$ nm laser pulses are applied in an iterative
way. We rely on the idea that when we apply the initialization sequence
$[$seg1$+$seg2$]$ as shown in Fig.\ref{fig:pulse_seq} for multiple
cycles, each subsequent cycle uses the final populations of a given
cycle as the initial state. Thus, when a cycle starts with an improved
purity of the target $\ket{0,0}$ state, the purity of the state is
enhanced further at the end of the cycle. We experimentally have implemented
the sequence $[$seg1 $+$ seg2$]$$\times N$ for $N=1$ to $5$
and have analyzed the resulting state using seg3. However, in this
way the resulting gain in the purity of the $\ket{0,0}$ state is
very small. An additional improvement can be obtained by adjusting
the durations of the laser pulses independently, using the rate equation
discussed in section III for each pulse depending on the relevant
initial states. Below we explain a protocol and demonstrate the results
of the simulation we have performed.

\begin{table}
\caption{Simulated values of the laser pulse duration and population of the
$\ket{0,0}$ state after each of the three cycles for seg1 and seg2}

\begin{tabular}{|>{\centering}p{0.14\columnwidth}|>{\centering}p{0.07\columnwidth}|>{\centering}p{0.14\columnwidth}|>{\centering}p{0.14\columnwidth}|>{\centering}p{0.14\columnwidth}|}
\hline 
Number of cycle & \multicolumn{1}{c|}{Duration of L$_{1}$(ns)} & $P_{\ket{0,0}}$ after seg1 (\%) & \multirow{1}{0.14\columnwidth}{Duration of L$_{2}$(ns)} & $P_{\ket{0,0}}$ after seg2 (\%)\tabularnewline
\hline 
\hline 
$1$ & $500$ & $55$ & $460$ & $70.9$\tabularnewline
\hline 
$2$ & $156$ & $72.3$ & $140$ & $73.5$\tabularnewline
\hline 
$3$ & $20$ & $74$ & $0$ & $74$\tabularnewline
\hline 
\end{tabular}\label{table simulation}
\end{table}

\begin{figure}
\includegraphics[width=1\columnwidth]{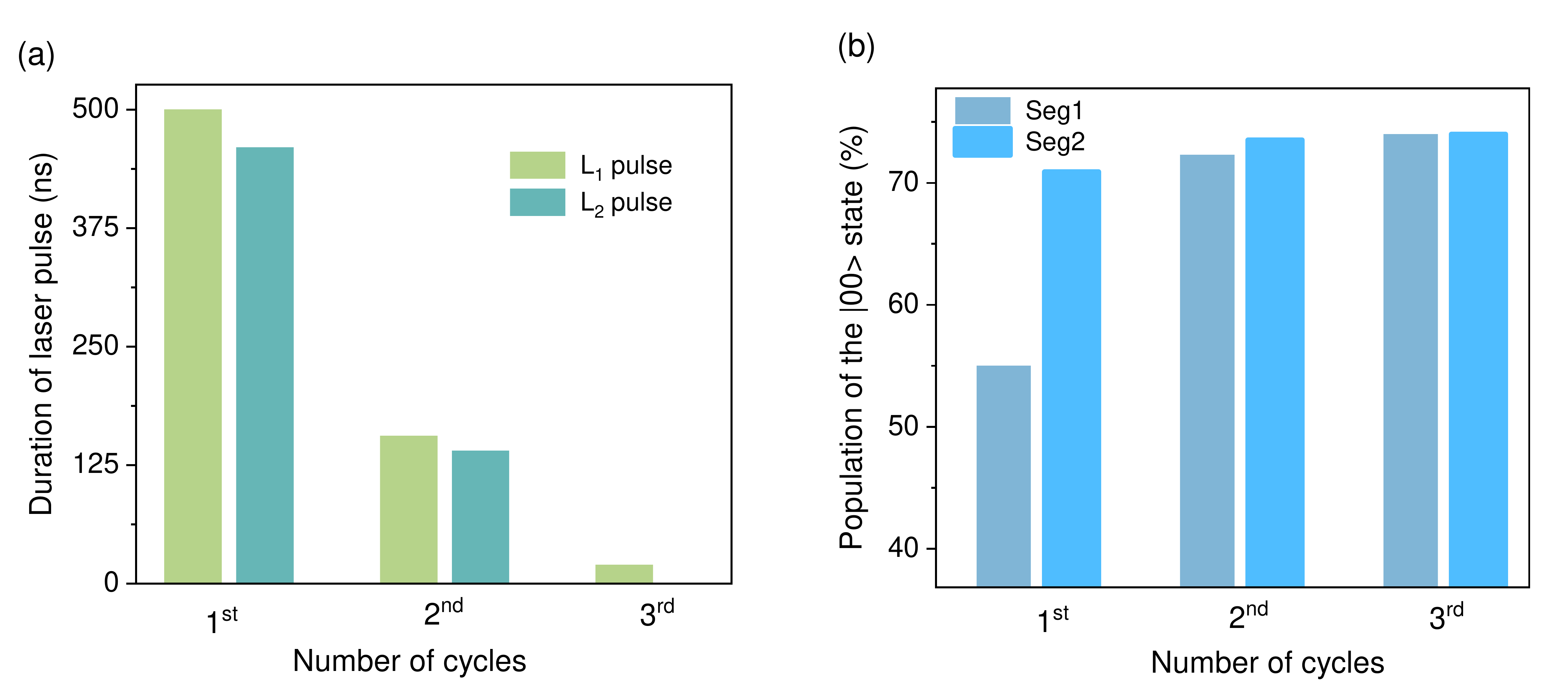}\caption{(a) Optimized duration of the laser pulses for three cycles for seg1
and (b) the resulting population of the $\ket{0,0}$ state after each
cycle. (c) Optimized laser pulse duration for seg2 and (d) purity
after each cycle. \label{fig:simulation}}
\end{figure}

We simulate the implementation of seg1 and seg2 in an iterative way:
we simulate the population redistribution for a sequence $[$seg1
$+$ seg2$]$$\times N$ where the duration of the pulses L$_{1}$
and L$_{2}$ are optimized for each cycle. For the $1^{st}$ cycle
we choose the duration of the L$_{1}$ and L$_{2}$ pulses as $500$
and $460$ ns as the resonance amplitude for the nuclear spin $m_{I}=0$
is maximum for this value {[}Fig.\ref{fig:populations_seg1}{]}. For
the $2^{nd}$ cycle, we consider the experimentally obtained final
population after $1^{st}$ cycle as the initial state and solve the
rate equation model for L$_{1}$and L$_{2}$ pulses. In this cycle,
we observe that maximum transfer of population to $\ket{0,0}$ state
occurs when the duration of the L$_{1}$and L$_{2}$ pulses are $156$
and $140$ ns. In this similar way, we solve the rate equation for
the $3^{rd}$ cycle. The optimized duration of the laser pulses and
the obtained purity $P_{\ket{0,0}}$ for three cycles are shown in
Fig.\ref{fig:simulation}(a) and (b). Table \ref{table simulation}
summarizes these values for both the segments. Moreover, we also have
simulated the population redistribution for a sequence $[$(seg1)$\times N$
$+$ (seg2)$\times N]$ where we first optimize the duration of the
laser pulses for three consecutive cycles of seg1 and then we do the
similar task for seg2. However, this method could not enhance the
purity than what we obtained using the sequence $[$seg1 $+$ seg2$]$$\times N$.

\section*{V. Discussions and conclusion}

The present paper demonstrates a detailed optimization procedure for
initializing a hybrid quantum spin register by performing the spin
manipulation in a reduced subspace of the spins. We apply a series
of MW and RF pulses to swap the electronic and nuclear spin polarization,
and laser pulses for optically pumping the electronic spin. The duration
of the laser pulses are optimized in such a way that we maximize the
electronic and nuclear spin polarization. For this purpose, we formulate
a rate equation model and investigate the polarization dynamics as
a function of the duration of the laser pulse. By performing partial
state tomography we experimentally determine the population of the
relevant eigenstates.

We follow an approach developed in ref. \citep{chakraborty2017polarizing}
to optimize the laser pulse duration. However, in this case we have
applied a different polarization method which is based on a population
trapping protocol. With this protocol, we access the subspace spanned
by $\ket{m_{S}=0,-1}\otimes\ket{m_{I}=0,\pm1}$ states and sequentially
populate the target state $\ket{0,0}$. For this purpose, a number
of MW, RF and optical pulses are employed to drive the population
from the initial state $\ket{0,\pm1}$ to the target state through
certain intermediate states in $m_{S}=-1$ subspace. We have optimized
the optical pumping processes to maximize the spin polarization. To
describe the present experimental results, we have used a rate equation
model which explains laser-induced redistribution of the population
among the six spin-levels. Furthermore, to reduce the nuclear depolarization,
we have formulated a recursive method where in each cycle the laser
pulses perform the optical pumping in an optimum way. Our simulation
shows that when multiple cycles of the control sequence, with each
cycle having an optimum laser pulse duration, are implemented, an
enhancement of the polarization can be achieved.

Initializing a quantum register to an eigenstate is important for
implementation of quantum information tasks. In these protocols, optical
control of the spins plays a significant role. The present results
of the optimization of the optical control pulses can help developing
spin manipulation protocols in deterministic way. One advantage of
our method is that the spin control can be performed in an arbitrary
magnetic field. We have developed the present method for an NV spin
register. However, this spin manipulation approach is general. Hence,
for other atomic defect centers in diamond which has optically addressable
spin transitions, a similar method can be formulated.
\begin{acknowledgments}
This work was supported by the DFG through grants 192/19-2, 192/34-1
and by the MERCUR foundation through grant Pr-2013-0003. 
\end{acknowledgments}

\section*{appendix}

During the optical pulse, the redistribution of the populations $\vec{P}$
can be described as

\begin{eqnarray*}
\frac{d\vec{P}}{dt} & = & M(k_{s},k_{I})\vec{P},
\end{eqnarray*}
where the matrix $M(k_{S},k_{I})$ of rate constants is

\[
M(k_{S},k_{I})=\left[\begin{array}{cccccc}
-2k_{I} & k_{I} & k_{I} & k_{S} & 0 & 0\\
k_{I} & -2k_{I} & k_{I} & 0 & k_{S} & 0\\
k_{I} & k_{I} & -2k_{I} & 0 & 0 & k_{S}\\
0 & 0 & 0 & -k_{S} & 0 & 0\\
0 & 0 & 0 & 0 & -k_{S} & 0\\
0 & 0 & 0 & 0 & 0 & -k_{S}
\end{array}\right]
\]
As discussed in the main text, the basis states are $(m_{s},m_{I})=(0,-1;0,+1;0,0;-1,-1;-1,+1;-1,0)$.
The constants $k_{I}$ and $k_{S}$ describe the rates at which the
electronic and nuclear spin polarisation are changed. Under our experimental
conditions, $1/k_{S}=0.27$ $\mu$s and $1/k_{I}=4.76$ $\mu$s \citep{chakraborty2017polarizing}.\\

In Seg1 the solution obtained for the initial condition $\vec{P}(0)=\frac{1}{3}(1,1,1,0,0,0)$
is

\begin{align*}
\vec{P}_{seg1}(t) & =\frac{1}{3}[1-\frac{k_{I}(e^{-k_{S}t}-e^{-3k_{I}t})}{(3k_{I}-k_{S})},\\
 & 1-\frac{(2k_{I}-k_{S})e^{-3k_{I}t}+k_{I}e^{-k_{S}t}}{(3k_{I}-k_{S})},\\
 & 1-\frac{(k_{I}-k_{S})e^{-k_{S}t}+(k_{s}-k_{I})e^{-3k_{I}t}}{(3k_{I}-k_{S})},\\
 & 0,0,e^{-k_{S}t}].\\
\end{align*}

For initial condition $\vec{P}=(0.07,0,0.55,0,0.05,0.33)$ in seg2,
the solution is

\begin{align*}
\vec{P}_{seg2}(t)= & [0.34+\frac{e^{-3k_{I}t}(0.26k_{S}-0.4k_{I})-0.38k_{I}e^{-k_{S}t}}{3k_{I}-k_{S}},\\
 & 0.34-\frac{e^{-k_{S}t}(0.38k_{I}-0.05k_{S})-e^{-3k_{I}t}(0.63k_{I}-0.29k_{S})}{3k_{I}-k_{S}},\\
 & 0.34+\frac{e^{-3k_{I}t}(1.03k_{I}-0.55k_{S})-e^{-k_{S}t}(0.38k_{I}-0.33k_{S})}{3k_{I}-k_{S}}]\\
0 & ,0.05e^{-k_{S}t},0.33e^{-k_{S}t}].
\end{align*}

\bibliographystyle{apsrev4-2}
\bibliography{main}

\begin{thebibliography}{30}%
\makeatletter
\providecommand \@ifxundefined [1]{%
 \@ifx{#1\undefined}
}%
\providecommand \@ifnum [1]{%
 \ifnum #1\expandafter \@firstoftwo
 \else \expandafter \@secondoftwo
 \fi
}%
\providecommand \@ifx [1]{%
 \ifx #1\expandafter \@firstoftwo
 \else \expandafter \@secondoftwo
 \fi
}%
\providecommand \natexlab [1]{#1}%
\providecommand \enquote  [1]{``#1''}%
\providecommand \bibnamefont  [1]{#1}%
\providecommand \bibfnamefont [1]{#1}%
\providecommand \citenamefont [1]{#1}%
\providecommand \href@noop [0]{\@secondoftwo}%
\providecommand \href [0]{\begingroup \@sanitize@url \@href}%
\providecommand \@href[1]{\@@startlink{#1}\@@href}%
\providecommand \@@href[1]{\endgroup#1\@@endlink}%
\providecommand \@sanitize@url [0]{\catcode `\\12\catcode `\$12\catcode
  `\&12\catcode `\#12\catcode `\^12\catcode `\_12\catcode `\%12\relax}%
\providecommand \@@startlink[1]{}%
\providecommand \@@endlink[0]{}%
\providecommand \url  [0]{\begingroup\@sanitize@url \@url }%
\providecommand \@url [1]{\endgroup\@href {#1}{\urlprefix }}%
\providecommand \urlprefix  [0]{URL }%
\providecommand \Eprint [0]{\href }%
\providecommand \doibase [0]{https://doi.org/}%
\providecommand \selectlanguage [0]{\@gobble}%
\providecommand \bibinfo  [0]{\@secondoftwo}%
\providecommand \bibfield  [0]{\@secondoftwo}%
\providecommand \translation [1]{[#1]}%
\providecommand \BibitemOpen [0]{}%
\providecommand \bibitemStop [0]{}%
\providecommand \bibitemNoStop [0]{.\EOS\space}%
\providecommand \EOS [0]{\spacefactor3000\relax}%
\providecommand \BibitemShut  [1]{\csname bibitem#1\endcsname}%
\let\auto@bib@innerbib\@empty
\bibitem [{\citenamefont {Jelezko}\ and\ \citenamefont
  {Wrachtrup}(2006)}]{Jelezko_review}%
  \BibitemOpen
  \bibfield  {author} {\bibinfo {author} {\bibfnamefont {F.}~\bibnamefont
  {Jelezko}}\ and\ \bibinfo {author} {\bibfnamefont {J.}~\bibnamefont
  {Wrachtrup}},\ }\href@noop {} {\bibfield  {journal} {\bibinfo  {journal}
  {Phys. Stat. Sol. (a)}\ }\textbf {\bibinfo {volume} {13}},\ \bibinfo {pages}
  {3207} (\bibinfo {year} {2006})}\BibitemShut {NoStop}%
\bibitem [{\citenamefont {Suter}\ and\ \citenamefont
  {Jelezko}(2017)}]{suter2017single}%
  \BibitemOpen
  \bibfield  {author} {\bibinfo {author} {\bibfnamefont {D.}~\bibnamefont
  {Suter}}\ and\ \bibinfo {author} {\bibfnamefont {F.}~\bibnamefont
  {Jelezko}},\ }\href@noop {} {\bibfield  {journal} {\bibinfo  {journal}
  {Progress in nuclear magnetic resonance spectroscopy}\ }\textbf {\bibinfo
  {volume} {98}},\ \bibinfo {pages} {50} (\bibinfo {year} {2017})}\BibitemShut
  {NoStop}%
\bibitem [{\citenamefont {Balasubramanian}\ \emph {et~al.}(2009)\citenamefont
  {Balasubramanian}, \citenamefont {Neumann}, \citenamefont {Twitchen},
  \citenamefont {Markham}, \citenamefont {Kolesov}, \citenamefont {Mizuochi},
  \citenamefont {Isoya}, \citenamefont {Achard}, \citenamefont {Beck},
  \citenamefont {Tissler} \emph {et~al.}}]{balasubramanian2009ultralong}%
  \BibitemOpen
  \bibfield  {author} {\bibinfo {author} {\bibfnamefont {G.}~\bibnamefont
  {Balasubramanian}}, \bibinfo {author} {\bibfnamefont {P.}~\bibnamefont
  {Neumann}}, \bibinfo {author} {\bibfnamefont {D.}~\bibnamefont {Twitchen}},
  \bibinfo {author} {\bibfnamefont {M.}~\bibnamefont {Markham}}, \bibinfo
  {author} {\bibfnamefont {R.}~\bibnamefont {Kolesov}}, \bibinfo {author}
  {\bibfnamefont {N.}~\bibnamefont {Mizuochi}}, \bibinfo {author}
  {\bibfnamefont {J.}~\bibnamefont {Isoya}}, \bibinfo {author} {\bibfnamefont
  {J.}~\bibnamefont {Achard}}, \bibinfo {author} {\bibfnamefont
  {J.}~\bibnamefont {Beck}}, \bibinfo {author} {\bibfnamefont {J.}~\bibnamefont
  {Tissler}}, \emph {et~al.},\ }\href@noop {} {\bibfield  {journal} {\bibinfo
  {journal} {Nature materials}\ }\textbf {\bibinfo {volume} {8}},\ \bibinfo
  {pages} {383} (\bibinfo {year} {2009})}\BibitemShut {NoStop}%
\bibitem [{\citenamefont {Kubanek}\ \emph {et~al.}(2012)\citenamefont
  {Kubanek}, \citenamefont {Sipahigil}, \citenamefont {Togan}, \citenamefont
  {Goldman}, \citenamefont {Chu}, \citenamefont {de~Leon}, \citenamefont
  {Zibrov},\ and\ \citenamefont {Lukin}}]{kubanek2012quantum}%
  \BibitemOpen
  \bibfield  {author} {\bibinfo {author} {\bibfnamefont {A.}~\bibnamefont
  {Kubanek}}, \bibinfo {author} {\bibfnamefont {A.}~\bibnamefont {Sipahigil}},
  \bibinfo {author} {\bibfnamefont {E.}~\bibnamefont {Togan}}, \bibinfo
  {author} {\bibfnamefont {M.}~\bibnamefont {Goldman}}, \bibinfo {author}
  {\bibfnamefont {Y.}~\bibnamefont {Chu}}, \bibinfo {author} {\bibfnamefont
  {N.}~\bibnamefont {de~Leon}}, \bibinfo {author} {\bibfnamefont
  {A.}~\bibnamefont {Zibrov}},\ and\ \bibinfo {author} {\bibfnamefont
  {M.}~\bibnamefont {Lukin}},\ }in\ \href@noop {} {\emph {\bibinfo {booktitle}
  {APS March Meeting Abstracts}}},\ Vol.\ \bibinfo {volume} {2012}\ (\bibinfo
  {year} {2012})\ pp.\ \bibinfo {pages} {D29--003}\BibitemShut {NoStop}%
\bibitem [{\citenamefont {Bernien}\ \emph {et~al.}(2012)\citenamefont
  {Bernien}, \citenamefont {Childress}, \citenamefont {Robledo}, \citenamefont
  {Markham}, \citenamefont {Twitchen},\ and\ \citenamefont
  {Hanson}}]{bernien2012two}%
  \BibitemOpen
  \bibfield  {author} {\bibinfo {author} {\bibfnamefont {H.}~\bibnamefont
  {Bernien}}, \bibinfo {author} {\bibfnamefont {L.}~\bibnamefont {Childress}},
  \bibinfo {author} {\bibfnamefont {L.}~\bibnamefont {Robledo}}, \bibinfo
  {author} {\bibfnamefont {M.}~\bibnamefont {Markham}}, \bibinfo {author}
  {\bibfnamefont {D.}~\bibnamefont {Twitchen}},\ and\ \bibinfo {author}
  {\bibfnamefont {R.}~\bibnamefont {Hanson}},\ }\href@noop {} {\bibfield
  {journal} {\bibinfo  {journal} {Physical Review Letters}\ }\textbf {\bibinfo
  {volume} {108}},\ \bibinfo {pages} {043604} (\bibinfo {year}
  {2012})}\BibitemShut {NoStop}%
\bibitem [{\citenamefont {Pezzagna}\ and\ \citenamefont
  {Meijer}(2021)}]{pezzagna2021quantum}%
  \BibitemOpen
  \bibfield  {author} {\bibinfo {author} {\bibfnamefont {S.}~\bibnamefont
  {Pezzagna}}\ and\ \bibinfo {author} {\bibfnamefont {J.}~\bibnamefont
  {Meijer}},\ }\href@noop {} {\bibfield  {journal} {\bibinfo  {journal}
  {Applied Physics Reviews}\ }\textbf {\bibinfo {volume} {8}},\ \bibinfo
  {pages} {011308} (\bibinfo {year} {2021})}\BibitemShut {NoStop}%
\bibitem [{\citenamefont {Childress}\ and\ \citenamefont
  {Hanson}(2013)}]{childress2013diamond}%
  \BibitemOpen
  \bibfield  {author} {\bibinfo {author} {\bibfnamefont {L.}~\bibnamefont
  {Childress}}\ and\ \bibinfo {author} {\bibfnamefont {R.}~\bibnamefont
  {Hanson}},\ }\href@noop {} {\bibfield  {journal} {\bibinfo  {journal} {MRS
  bulletin}\ }\textbf {\bibinfo {volume} {38}},\ \bibinfo {pages} {134}
  (\bibinfo {year} {2013})}\BibitemShut {NoStop}%
\bibitem [{\citenamefont {Fabbri}\ and\ \citenamefont
  {Hern{\'a}ndez-G{\'o}mez}(2020)}]{fabbri2020quantum}%
  \BibitemOpen
  \bibfield  {author} {\bibinfo {author} {\bibfnamefont {N.}~\bibnamefont
  {Fabbri}}\ and\ \bibinfo {author} {\bibfnamefont {S.}~\bibnamefont
  {Hern{\'a}ndez-G{\'o}mez}},\ }\href@noop {} {\bibfield  {journal} {\bibinfo
  {journal} {Frontiers in Physics}\ }\textbf {\bibinfo {volume} {8}},\ \bibinfo
  {pages} {652} (\bibinfo {year} {2020})}\BibitemShut {NoStop}%
\bibitem [{\citenamefont {Vasconcelos}\ \emph {et~al.}(2020)\citenamefont
  {Vasconcelos}, \citenamefont {Reisenbauer}, \citenamefont {Salter},
  \citenamefont {Wachter}, \citenamefont {Wirtitsch}, \citenamefont
  {Schmiedmayer}, \citenamefont {Walther},\ and\ \citenamefont
  {Trupke}}]{vasconcelos2020scalable}%
  \BibitemOpen
  \bibfield  {author} {\bibinfo {author} {\bibfnamefont {R.}~\bibnamefont
  {Vasconcelos}}, \bibinfo {author} {\bibfnamefont {S.}~\bibnamefont
  {Reisenbauer}}, \bibinfo {author} {\bibfnamefont {C.}~\bibnamefont {Salter}},
  \bibinfo {author} {\bibfnamefont {G.}~\bibnamefont {Wachter}}, \bibinfo
  {author} {\bibfnamefont {D.}~\bibnamefont {Wirtitsch}}, \bibinfo {author}
  {\bibfnamefont {J.}~\bibnamefont {Schmiedmayer}}, \bibinfo {author}
  {\bibfnamefont {P.}~\bibnamefont {Walther}},\ and\ \bibinfo {author}
  {\bibfnamefont {M.}~\bibnamefont {Trupke}},\ }\href@noop {} {\bibfield
  {journal} {\bibinfo  {journal} {npj Quantum Information}\ }\textbf {\bibinfo
  {volume} {6}},\ \bibinfo {pages} {1} (\bibinfo {year} {2020})}\BibitemShut
  {NoStop}%
\bibitem [{\citenamefont {Pfaff}\ \emph {et~al.}(2014)\citenamefont {Pfaff},
  \citenamefont {Hensen}, \citenamefont {Bernien}, \citenamefont {van Dam},
  \citenamefont {Blok}, \citenamefont {Taminiau}, \citenamefont {Tiggelman},
  \citenamefont {Schouten}, \citenamefont {Markham}, \citenamefont {Twitchen}
  \emph {et~al.}}]{pfaff2014unconditional}%
  \BibitemOpen
  \bibfield  {author} {\bibinfo {author} {\bibfnamefont {W.}~\bibnamefont
  {Pfaff}}, \bibinfo {author} {\bibfnamefont {B.~J.}\ \bibnamefont {Hensen}},
  \bibinfo {author} {\bibfnamefont {H.}~\bibnamefont {Bernien}}, \bibinfo
  {author} {\bibfnamefont {S.~B.}\ \bibnamefont {van Dam}}, \bibinfo {author}
  {\bibfnamefont {M.~S.}\ \bibnamefont {Blok}}, \bibinfo {author}
  {\bibfnamefont {T.~H.}\ \bibnamefont {Taminiau}}, \bibinfo {author}
  {\bibfnamefont {M.~J.}\ \bibnamefont {Tiggelman}}, \bibinfo {author}
  {\bibfnamefont {R.~N.}\ \bibnamefont {Schouten}}, \bibinfo {author}
  {\bibfnamefont {M.}~\bibnamefont {Markham}}, \bibinfo {author} {\bibfnamefont
  {D.~J.}\ \bibnamefont {Twitchen}}, \emph {et~al.},\ }\href@noop {} {\bibfield
   {journal} {\bibinfo  {journal} {Science}\ }\textbf {\bibinfo {volume}
  {345}},\ \bibinfo {pages} {532} (\bibinfo {year} {2014})}\BibitemShut
  {NoStop}%
\bibitem [{\citenamefont {Zhang}\ \emph {et~al.}(2020)\citenamefont {Zhang},
  \citenamefont {Hegde},\ and\ \citenamefont {Suter}}]{zhang2020efficient}%
  \BibitemOpen
  \bibfield  {author} {\bibinfo {author} {\bibfnamefont {J.}~\bibnamefont
  {Zhang}}, \bibinfo {author} {\bibfnamefont {S.~S.}\ \bibnamefont {Hegde}},\
  and\ \bibinfo {author} {\bibfnamefont {D.}~\bibnamefont {Suter}},\
  }\href@noop {} {\bibfield  {journal} {\bibinfo  {journal} {Physical Review
  Letters}\ }\textbf {\bibinfo {volume} {125}},\ \bibinfo {pages} {030501}
  (\bibinfo {year} {2020})}\BibitemShut {NoStop}%
\bibitem [{\citenamefont {Pompili}\ \emph {et~al.}(2021)\citenamefont
  {Pompili}, \citenamefont {Hermans}, \citenamefont {Baier}, \citenamefont
  {Beukers}, \citenamefont {Humphreys}, \citenamefont {Schouten}, \citenamefont
  {Vermeulen}, \citenamefont {Tiggelman}, \citenamefont {Martins},
  \citenamefont {Dirkse} \emph {et~al.}}]{pompili2021realization}%
  \BibitemOpen
  \bibfield  {author} {\bibinfo {author} {\bibfnamefont {M.}~\bibnamefont
  {Pompili}}, \bibinfo {author} {\bibfnamefont {S.~L.}\ \bibnamefont
  {Hermans}}, \bibinfo {author} {\bibfnamefont {S.}~\bibnamefont {Baier}},
  \bibinfo {author} {\bibfnamefont {H.~K.}\ \bibnamefont {Beukers}}, \bibinfo
  {author} {\bibfnamefont {P.~C.}\ \bibnamefont {Humphreys}}, \bibinfo {author}
  {\bibfnamefont {R.~N.}\ \bibnamefont {Schouten}}, \bibinfo {author}
  {\bibfnamefont {R.~F.}\ \bibnamefont {Vermeulen}}, \bibinfo {author}
  {\bibfnamefont {M.~J.}\ \bibnamefont {Tiggelman}}, \bibinfo {author}
  {\bibfnamefont {L.~d.~S.}\ \bibnamefont {Martins}}, \bibinfo {author}
  {\bibfnamefont {B.}~\bibnamefont {Dirkse}}, \emph {et~al.},\ }\href@noop {}
  {\bibfield  {journal} {\bibinfo  {journal} {arXiv preprint arXiv:2102.04471}\
  } (\bibinfo {year} {2021})}\BibitemShut {NoStop}%
\bibitem [{\citenamefont {Nemoto}\ \emph {et~al.}(2016)\citenamefont {Nemoto},
  \citenamefont {Trupke}, \citenamefont {Devitt}, \citenamefont
  {Scharfenberger}, \citenamefont {Buczak}, \citenamefont {Schmiedmayer},\ and\
  \citenamefont {Munro}}]{nemoto2016photonic}%
  \BibitemOpen
  \bibfield  {author} {\bibinfo {author} {\bibfnamefont {K.}~\bibnamefont
  {Nemoto}}, \bibinfo {author} {\bibfnamefont {M.}~\bibnamefont {Trupke}},
  \bibinfo {author} {\bibfnamefont {S.~J.}\ \bibnamefont {Devitt}}, \bibinfo
  {author} {\bibfnamefont {B.}~\bibnamefont {Scharfenberger}}, \bibinfo
  {author} {\bibfnamefont {K.}~\bibnamefont {Buczak}}, \bibinfo {author}
  {\bibfnamefont {J.}~\bibnamefont {Schmiedmayer}},\ and\ \bibinfo {author}
  {\bibfnamefont {W.~J.}\ \bibnamefont {Munro}},\ }\href@noop {} {\bibfield
  {journal} {\bibinfo  {journal} {Scientific reports}\ }\textbf {\bibinfo
  {volume} {6}},\ \bibinfo {pages} {1} (\bibinfo {year} {2016})}\BibitemShut
  {NoStop}%
\bibitem [{\citenamefont {Degen}\ \emph {et~al.}(2017)\citenamefont {Degen},
  \citenamefont {Reinhard},\ and\ \citenamefont
  {Cappellaro}}]{degen2017quantum}%
  \BibitemOpen
  \bibfield  {author} {\bibinfo {author} {\bibfnamefont {C.~L.}\ \bibnamefont
  {Degen}}, \bibinfo {author} {\bibfnamefont {F.}~\bibnamefont {Reinhard}},\
  and\ \bibinfo {author} {\bibfnamefont {P.}~\bibnamefont {Cappellaro}},\
  }\href@noop {} {\bibfield  {journal} {\bibinfo  {journal} {Reviews of modern
  physics}\ }\textbf {\bibinfo {volume} {89}},\ \bibinfo {pages} {035002}
  (\bibinfo {year} {2017})}\BibitemShut {NoStop}%
\bibitem [{\citenamefont {Schirhagl}\ \emph {et~al.}(2014)\citenamefont
  {Schirhagl}, \citenamefont {Chang}, \citenamefont {Loretz},\ and\
  \citenamefont {Degen}}]{schirhagl2014nitrogen}%
  \BibitemOpen
  \bibfield  {author} {\bibinfo {author} {\bibfnamefont {R.}~\bibnamefont
  {Schirhagl}}, \bibinfo {author} {\bibfnamefont {K.}~\bibnamefont {Chang}},
  \bibinfo {author} {\bibfnamefont {M.}~\bibnamefont {Loretz}},\ and\ \bibinfo
  {author} {\bibfnamefont {C.~L.}\ \bibnamefont {Degen}},\ }\href@noop {}
  {\bibfield  {journal} {\bibinfo  {journal} {Annual review of physical
  chemistry}\ }\textbf {\bibinfo {volume} {65}},\ \bibinfo {pages} {83}
  (\bibinfo {year} {2014})}\BibitemShut {NoStop}%
\bibitem [{\citenamefont {Larsson}\ and\ \citenamefont
  {Delaney}(2008)}]{larsson2008electronic}%
  \BibitemOpen
  \bibfield  {author} {\bibinfo {author} {\bibfnamefont {J.}~\bibnamefont
  {Larsson}}\ and\ \bibinfo {author} {\bibfnamefont {P.}~\bibnamefont
  {Delaney}},\ }\href@noop {} {\bibfield  {journal} {\bibinfo  {journal}
  {Physical Review B}\ }\textbf {\bibinfo {volume} {77}},\ \bibinfo {pages}
  {165201} (\bibinfo {year} {2008})}\BibitemShut {NoStop}%
\bibitem [{\citenamefont {Doherty}\ \emph {et~al.}(2012)\citenamefont
  {Doherty}, \citenamefont {Dolde}, \citenamefont {Fedder}, \citenamefont
  {Jelezko}, \citenamefont {Wrachtrup}, \citenamefont {Manson},\ and\
  \citenamefont {Hollenberg}}]{doherty2012theory}%
  \BibitemOpen
  \bibfield  {author} {\bibinfo {author} {\bibfnamefont {M.}~\bibnamefont
  {Doherty}}, \bibinfo {author} {\bibfnamefont {F.}~\bibnamefont {Dolde}},
  \bibinfo {author} {\bibfnamefont {H.}~\bibnamefont {Fedder}}, \bibinfo
  {author} {\bibfnamefont {F.}~\bibnamefont {Jelezko}}, \bibinfo {author}
  {\bibfnamefont {J.}~\bibnamefont {Wrachtrup}}, \bibinfo {author}
  {\bibfnamefont {N.}~\bibnamefont {Manson}},\ and\ \bibinfo {author}
  {\bibfnamefont {L.}~\bibnamefont {Hollenberg}},\ }\href@noop {} {\bibfield
  {journal} {\bibinfo  {journal} {Physical Review B}\ }\textbf {\bibinfo
  {volume} {85}},\ \bibinfo {pages} {205203} (\bibinfo {year}
  {2012})}\BibitemShut {NoStop}%
\bibitem [{\citenamefont {Chakraborty}\ \emph {et~al.}(2019)\citenamefont
  {Chakraborty}, \citenamefont {Lehmann}, \citenamefont {Zhang}, \citenamefont
  {Borgsdorf}, \citenamefont {W{\"o}hrl}, \citenamefont {Remfort},
  \citenamefont {Buck}, \citenamefont {K{\"o}hler},\ and\ \citenamefont
  {Suter}}]{chakraborty2019cvd}%
  \BibitemOpen
  \bibfield  {author} {\bibinfo {author} {\bibfnamefont {T.}~\bibnamefont
  {Chakraborty}}, \bibinfo {author} {\bibfnamefont {F.}~\bibnamefont
  {Lehmann}}, \bibinfo {author} {\bibfnamefont {J.}~\bibnamefont {Zhang}},
  \bibinfo {author} {\bibfnamefont {S.}~\bibnamefont {Borgsdorf}}, \bibinfo
  {author} {\bibfnamefont {N.}~\bibnamefont {W{\"o}hrl}}, \bibinfo {author}
  {\bibfnamefont {R.}~\bibnamefont {Remfort}}, \bibinfo {author} {\bibfnamefont
  {V.}~\bibnamefont {Buck}}, \bibinfo {author} {\bibfnamefont {U.}~\bibnamefont
  {K{\"o}hler}},\ and\ \bibinfo {author} {\bibfnamefont {D.}~\bibnamefont
  {Suter}},\ }\href@noop {} {\bibfield  {journal} {\bibinfo  {journal}
  {Physical Review Materials}\ }\textbf {\bibinfo {volume} {3}},\ \bibinfo
  {pages} {065205} (\bibinfo {year} {2019})}\BibitemShut {NoStop}%
\bibitem [{\citenamefont {Dutt}\ \emph {et~al.}(2007)\citenamefont {Dutt},
  \citenamefont {Childress}, \citenamefont {Jiang}, \citenamefont {Togan},
  \citenamefont {Maze}, \citenamefont {Jelezko}, \citenamefont {Zibrov},
  \citenamefont {Hemmer},\ and\ \citenamefont {Lukin}}]{Dutta_Science}%
  \BibitemOpen
  \bibfield  {author} {\bibinfo {author} {\bibfnamefont {M.~V.~G.}\
  \bibnamefont {Dutt}}, \bibinfo {author} {\bibfnamefont {L.}~\bibnamefont
  {Childress}}, \bibinfo {author} {\bibfnamefont {L.}~\bibnamefont {Jiang}},
  \bibinfo {author} {\bibfnamefont {E.}~\bibnamefont {Togan}}, \bibinfo
  {author} {\bibfnamefont {J.}~\bibnamefont {Maze}}, \bibinfo {author}
  {\bibfnamefont {F.}~\bibnamefont {Jelezko}}, \bibinfo {author} {\bibfnamefont
  {A.~S.}\ \bibnamefont {Zibrov}}, \bibinfo {author} {\bibfnamefont {P.~R.}\
  \bibnamefont {Hemmer}},\ and\ \bibinfo {author} {\bibfnamefont {M.~D.}\
  \bibnamefont {Lukin}},\ }\href@noop {} {\bibfield  {journal} {\bibinfo
  {journal} {Phys. Rev. Lett.}\ }\textbf {\bibinfo {volume} {316}},\ \bibinfo
  {pages} {1312} (\bibinfo {year} {2007})}\BibitemShut {NoStop}%
\bibitem [{\citenamefont {Childress}\ \emph {et~al.}(2006)\citenamefont
  {Childress}, \citenamefont {Dutt}, \citenamefont {Taylor}, \citenamefont
  {Zibrov}, \citenamefont {Jelezko}, \citenamefont {Wrachtrup}, \citenamefont
  {Hemmer},\ and\ \citenamefont {Lukin}}]{childress2006coherent}%
  \BibitemOpen
  \bibfield  {author} {\bibinfo {author} {\bibfnamefont {L.}~\bibnamefont
  {Childress}}, \bibinfo {author} {\bibfnamefont {M.~G.}\ \bibnamefont {Dutt}},
  \bibinfo {author} {\bibfnamefont {J.}~\bibnamefont {Taylor}}, \bibinfo
  {author} {\bibfnamefont {A.}~\bibnamefont {Zibrov}}, \bibinfo {author}
  {\bibfnamefont {F.}~\bibnamefont {Jelezko}}, \bibinfo {author} {\bibfnamefont
  {J.}~\bibnamefont {Wrachtrup}}, \bibinfo {author} {\bibfnamefont
  {P.}~\bibnamefont {Hemmer}},\ and\ \bibinfo {author} {\bibfnamefont
  {M.}~\bibnamefont {Lukin}},\ }\href@noop {} {\bibfield  {journal} {\bibinfo
  {journal} {Science}\ }\textbf {\bibinfo {volume} {314}},\ \bibinfo {pages}
  {281} (\bibinfo {year} {2006})}\BibitemShut {NoStop}%
\bibitem [{\citenamefont {Gaebel}\ \emph {et~al.}(2006)\citenamefont {Gaebel},
  \citenamefont {Domhan}, \citenamefont {Popa}, \citenamefont {Wittmann},
  \citenamefont {Neumann}, \citenamefont {Jelezko}, \citenamefont {Rabeau},
  \citenamefont {Stavrias}, \citenamefont {Greentree}, \citenamefont {Prawer}
  \emph {et~al.}}]{gaebel2006room}%
  \BibitemOpen
  \bibfield  {author} {\bibinfo {author} {\bibfnamefont {T.}~\bibnamefont
  {Gaebel}}, \bibinfo {author} {\bibfnamefont {M.}~\bibnamefont {Domhan}},
  \bibinfo {author} {\bibfnamefont {I.}~\bibnamefont {Popa}}, \bibinfo {author}
  {\bibfnamefont {C.}~\bibnamefont {Wittmann}}, \bibinfo {author}
  {\bibfnamefont {P.}~\bibnamefont {Neumann}}, \bibinfo {author} {\bibfnamefont
  {F.}~\bibnamefont {Jelezko}}, \bibinfo {author} {\bibfnamefont {J.~R.}\
  \bibnamefont {Rabeau}}, \bibinfo {author} {\bibfnamefont {N.}~\bibnamefont
  {Stavrias}}, \bibinfo {author} {\bibfnamefont {A.~D.}\ \bibnamefont
  {Greentree}}, \bibinfo {author} {\bibfnamefont {S.}~\bibnamefont {Prawer}},
  \emph {et~al.},\ }\href@noop {} {\bibfield  {journal} {\bibinfo  {journal}
  {Nature Physics}\ }\textbf {\bibinfo {volume} {2}},\ \bibinfo {pages} {408}
  (\bibinfo {year} {2006})}\BibitemShut {NoStop}%
\bibitem [{\citenamefont {Chakraborty}\ \emph {et~al.}(2017)\citenamefont
  {Chakraborty}, \citenamefont {Zhang},\ and\ \citenamefont
  {Suter}}]{chakraborty2017polarizing}%
  \BibitemOpen
  \bibfield  {author} {\bibinfo {author} {\bibfnamefont {T.}~\bibnamefont
  {Chakraborty}}, \bibinfo {author} {\bibfnamefont {J.}~\bibnamefont {Zhang}},\
  and\ \bibinfo {author} {\bibfnamefont {D.}~\bibnamefont {Suter}},\
  }\href@noop {} {\bibfield  {journal} {\bibinfo  {journal} {New Journal of
  Physics}\ }\textbf {\bibinfo {volume} {19}},\ \bibinfo {pages} {073030}
  (\bibinfo {year} {2017})}\BibitemShut {NoStop}%
\bibitem [{\citenamefont {Pagliero}\ \emph {et~al.}(2014)\citenamefont
  {Pagliero}, \citenamefont {Laraoui}, \citenamefont {Henshaw},\ and\
  \citenamefont {Meriles}}]{APL(105)242402}%
  \BibitemOpen
  \bibfield  {author} {\bibinfo {author} {\bibfnamefont {D.}~\bibnamefont
  {Pagliero}}, \bibinfo {author} {\bibfnamefont {A.}~\bibnamefont {Laraoui}},
  \bibinfo {author} {\bibfnamefont {J.~D.}\ \bibnamefont {Henshaw}},\ and\
  \bibinfo {author} {\bibfnamefont {C.~A.}\ \bibnamefont {Meriles}},\
  }\href@noop {} {\bibfield  {journal} {\bibinfo  {journal} {Applied Physics
  Letters}\ }\textbf {\bibinfo {volume} {105}},\ \bibinfo {pages} {242402}
  (\bibinfo {year} {2014})}\BibitemShut {NoStop}%
\bibitem [{\citenamefont {Rao}\ \emph {et~al.}(2020)\citenamefont {Rao},
  \citenamefont {Wang}, \citenamefont {Zhang},\ and\ \citenamefont
  {Suter}}]{rao2020optimal}%
  \BibitemOpen
  \bibfield  {author} {\bibinfo {author} {\bibfnamefont {K.~R.~K.}\
  \bibnamefont {Rao}}, \bibinfo {author} {\bibfnamefont {Y.}~\bibnamefont
  {Wang}}, \bibinfo {author} {\bibfnamefont {J.}~\bibnamefont {Zhang}},\ and\
  \bibinfo {author} {\bibfnamefont {D.}~\bibnamefont {Suter}},\ }\href@noop {}
  {\bibfield  {journal} {\bibinfo  {journal} {Physical Review A}\ }\textbf
  {\bibinfo {volume} {101}},\ \bibinfo {pages} {013835} (\bibinfo {year}
  {2020})}\BibitemShut {NoStop}%
\bibitem [{\citenamefont {Jacques}\ \emph {et~al.}(2009)\citenamefont
  {Jacques}, \citenamefont {Neumann}, \citenamefont {Beck}, \citenamefont
  {Markham}, \citenamefont {Twitchen}, \citenamefont {Meijer}, \citenamefont
  {Kaiser}, \citenamefont {Balasubramanian}, \citenamefont {Jelezko},\ and\
  \citenamefont {Wrachtrup}}]{jacques2009dynamic}%
  \BibitemOpen
  \bibfield  {author} {\bibinfo {author} {\bibfnamefont {V.}~\bibnamefont
  {Jacques}}, \bibinfo {author} {\bibfnamefont {P.}~\bibnamefont {Neumann}},
  \bibinfo {author} {\bibfnamefont {J.}~\bibnamefont {Beck}}, \bibinfo {author}
  {\bibfnamefont {M.}~\bibnamefont {Markham}}, \bibinfo {author} {\bibfnamefont
  {D.}~\bibnamefont {Twitchen}}, \bibinfo {author} {\bibfnamefont
  {J.}~\bibnamefont {Meijer}}, \bibinfo {author} {\bibfnamefont
  {F.}~\bibnamefont {Kaiser}}, \bibinfo {author} {\bibfnamefont
  {G.}~\bibnamefont {Balasubramanian}}, \bibinfo {author} {\bibfnamefont
  {F.}~\bibnamefont {Jelezko}},\ and\ \bibinfo {author} {\bibfnamefont
  {J.}~\bibnamefont {Wrachtrup}},\ }\href@noop {} {\bibfield  {journal}
  {\bibinfo  {journal} {Physical review letters}\ }\textbf {\bibinfo {volume}
  {102}},\ \bibinfo {pages} {057403} (\bibinfo {year} {2009})}\BibitemShut
  {NoStop}%
\bibitem [{\citenamefont {{\'A}lvarez}\ \emph {et~al.}(2015)\citenamefont
  {{\'A}lvarez}, \citenamefont {Bretschneider}, \citenamefont {Fischer},
  \citenamefont {London}, \citenamefont {Kanda}, \citenamefont {Onoda},
  \citenamefont {Isoya}, \citenamefont {Gershoni},\ and\ \citenamefont
  {Frydman}}]{alvarez2015local}%
  \BibitemOpen
  \bibfield  {author} {\bibinfo {author} {\bibfnamefont {G.~A.}\ \bibnamefont
  {{\'A}lvarez}}, \bibinfo {author} {\bibfnamefont {C.~O.}\ \bibnamefont
  {Bretschneider}}, \bibinfo {author} {\bibfnamefont {R.}~\bibnamefont
  {Fischer}}, \bibinfo {author} {\bibfnamefont {P.}~\bibnamefont {London}},
  \bibinfo {author} {\bibfnamefont {H.}~\bibnamefont {Kanda}}, \bibinfo
  {author} {\bibfnamefont {S.}~\bibnamefont {Onoda}}, \bibinfo {author}
  {\bibfnamefont {J.}~\bibnamefont {Isoya}}, \bibinfo {author} {\bibfnamefont
  {D.}~\bibnamefont {Gershoni}},\ and\ \bibinfo {author} {\bibfnamefont
  {L.}~\bibnamefont {Frydman}},\ }\href@noop {} {\bibfield  {journal} {\bibinfo
   {journal} {Nature communications}\ }\textbf {\bibinfo {volume} {6}},\
  \bibinfo {pages} {1} (\bibinfo {year} {2015})}\BibitemShut {NoStop}%
\bibitem [{\citenamefont {Scheuer}\ \emph {et~al.}(2016)\citenamefont
  {Scheuer}, \citenamefont {Schwartz}, \citenamefont {Chen}, \citenamefont
  {Schulze-S{\"u}nninghausen}, \citenamefont {Carl}, \citenamefont {H{\"o}fer},
  \citenamefont {Retzker}, \citenamefont {Sumiya}, \citenamefont {Isoya},
  \citenamefont {Luy} \emph {et~al.}}]{scheuer2016optically}%
  \BibitemOpen
  \bibfield  {author} {\bibinfo {author} {\bibfnamefont {J.}~\bibnamefont
  {Scheuer}}, \bibinfo {author} {\bibfnamefont {I.}~\bibnamefont {Schwartz}},
  \bibinfo {author} {\bibfnamefont {Q.}~\bibnamefont {Chen}}, \bibinfo {author}
  {\bibfnamefont {D.}~\bibnamefont {Schulze-S{\"u}nninghausen}}, \bibinfo
  {author} {\bibfnamefont {P.}~\bibnamefont {Carl}}, \bibinfo {author}
  {\bibfnamefont {P.}~\bibnamefont {H{\"o}fer}}, \bibinfo {author}
  {\bibfnamefont {A.}~\bibnamefont {Retzker}}, \bibinfo {author} {\bibfnamefont
  {H.}~\bibnamefont {Sumiya}}, \bibinfo {author} {\bibfnamefont
  {J.}~\bibnamefont {Isoya}}, \bibinfo {author} {\bibfnamefont
  {B.}~\bibnamefont {Luy}}, \emph {et~al.},\ }\href@noop {} {\bibfield
  {journal} {\bibinfo  {journal} {New journal of Physics}\ }\textbf {\bibinfo
  {volume} {18}},\ \bibinfo {pages} {013040} (\bibinfo {year}
  {2016})}\BibitemShut {NoStop}%
\bibitem [{\citenamefont {King}\ \emph {et~al.}(2015)\citenamefont {King},
  \citenamefont {Jeong}, \citenamefont {Vassiliou}, \citenamefont {Shin},
  \citenamefont {Page}, \citenamefont {Avalos}, \citenamefont {Wang},\ and\
  \citenamefont {Pines}}]{king2015room}%
  \BibitemOpen
  \bibfield  {author} {\bibinfo {author} {\bibfnamefont {J.~P.}\ \bibnamefont
  {King}}, \bibinfo {author} {\bibfnamefont {K.}~\bibnamefont {Jeong}},
  \bibinfo {author} {\bibfnamefont {C.~C.}\ \bibnamefont {Vassiliou}}, \bibinfo
  {author} {\bibfnamefont {C.~S.}\ \bibnamefont {Shin}}, \bibinfo {author}
  {\bibfnamefont {R.~H.}\ \bibnamefont {Page}}, \bibinfo {author}
  {\bibfnamefont {C.~E.}\ \bibnamefont {Avalos}}, \bibinfo {author}
  {\bibfnamefont {H.-J.}\ \bibnamefont {Wang}},\ and\ \bibinfo {author}
  {\bibfnamefont {A.}~\bibnamefont {Pines}},\ }\href@noop {} {\bibfield
  {journal} {\bibinfo  {journal} {Nature communications}\ }\textbf {\bibinfo
  {volume} {6}},\ \bibinfo {pages} {1} (\bibinfo {year} {2015})}\BibitemShut
  {NoStop}%
\bibitem [{\citenamefont {Jamonneau}\ \emph {et~al.}(2016)\citenamefont
  {Jamonneau}, \citenamefont {H{\'e}tet}, \citenamefont {Dr{\'e}au},
  \citenamefont {Roch},\ and\ \citenamefont {Jacques}}]{jamonneau2016coherent}%
  \BibitemOpen
  \bibfield  {author} {\bibinfo {author} {\bibfnamefont {P.}~\bibnamefont
  {Jamonneau}}, \bibinfo {author} {\bibfnamefont {G.}~\bibnamefont
  {H{\'e}tet}}, \bibinfo {author} {\bibfnamefont {A.}~\bibnamefont
  {Dr{\'e}au}}, \bibinfo {author} {\bibfnamefont {J.-F.}\ \bibnamefont
  {Roch}},\ and\ \bibinfo {author} {\bibfnamefont {V.}~\bibnamefont
  {Jacques}},\ }\href@noop {} {\bibfield  {journal} {\bibinfo  {journal}
  {Physical review letters}\ }\textbf {\bibinfo {volume} {116}},\ \bibinfo
  {pages} {043603} (\bibinfo {year} {2016})}\BibitemShut {NoStop}%
\bibitem [{\citenamefont {Nicolas}\ \emph {et~al.}(2018)\citenamefont
  {Nicolas}, \citenamefont {Delord}, \citenamefont {Jamonneau}, \citenamefont
  {Coto}, \citenamefont {Maze}, \citenamefont {Jacques},\ and\ \citenamefont
  {H{\'e}tet}}]{nicolas2018coherent}%
  \BibitemOpen
  \bibfield  {author} {\bibinfo {author} {\bibfnamefont {L.}~\bibnamefont
  {Nicolas}}, \bibinfo {author} {\bibfnamefont {T.}~\bibnamefont {Delord}},
  \bibinfo {author} {\bibfnamefont {P.}~\bibnamefont {Jamonneau}}, \bibinfo
  {author} {\bibfnamefont {R.}~\bibnamefont {Coto}}, \bibinfo {author}
  {\bibfnamefont {J.}~\bibnamefont {Maze}}, \bibinfo {author} {\bibfnamefont
  {V.}~\bibnamefont {Jacques}},\ and\ \bibinfo {author} {\bibfnamefont
  {G.}~\bibnamefont {H{\'e}tet}},\ }\href@noop {} {\bibfield  {journal}
  {\bibinfo  {journal} {New Journal of Physics}\ }\textbf {\bibinfo {volume}
  {20}},\ \bibinfo {pages} {033007} (\bibinfo {year} {2018})}\BibitemShut
  {NoStop}%
\end{thebibliography}%

\end{document}